\documentclass[electronic]{vgtc}             
 
\graphicspath{{figures/}{pictures/}{images/}{./}} 
\usepackage{enumitem}

\usepackage{times}                     

\usepackage{xkeyval}

\usepackage{graphicx}

\usepackage{booktabs}
\usepackage{mathptmx}                  

\vgtccategory{Research}

\vgtcinsertpkg

\title{From Voices to Worlds: Developing an AI-Powered Framework for 3D Object Generation in Augmented Reality}

\author{Majid Behravan\thanks{e-mail: behravan@vt.edu}\\ %
\parbox{2in}{\scriptsize \centering Department of Computer Science \\ Virginia Tech}
\and Denis Gra{\v{c}}anin\thanks{e-mail: gracanin@vt.edu}\\ %
\parbox{2in}{\scriptsize \centering Department of Computer Science \\ Virginia Tech}}

\teaser{
  \centering
\includegraphics[width=0.3\textwidth]{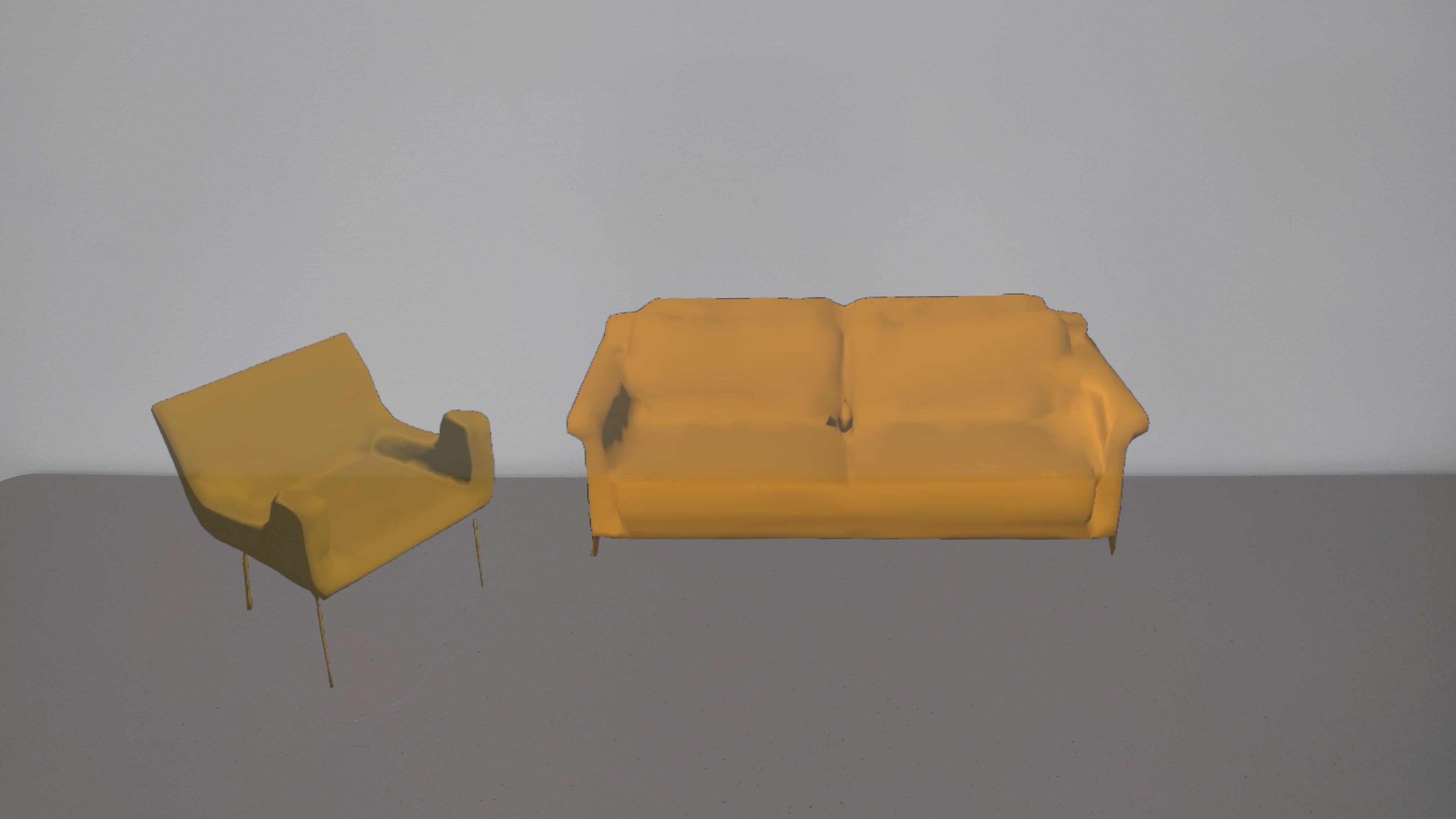}
\includegraphics[width=0.3\textwidth]{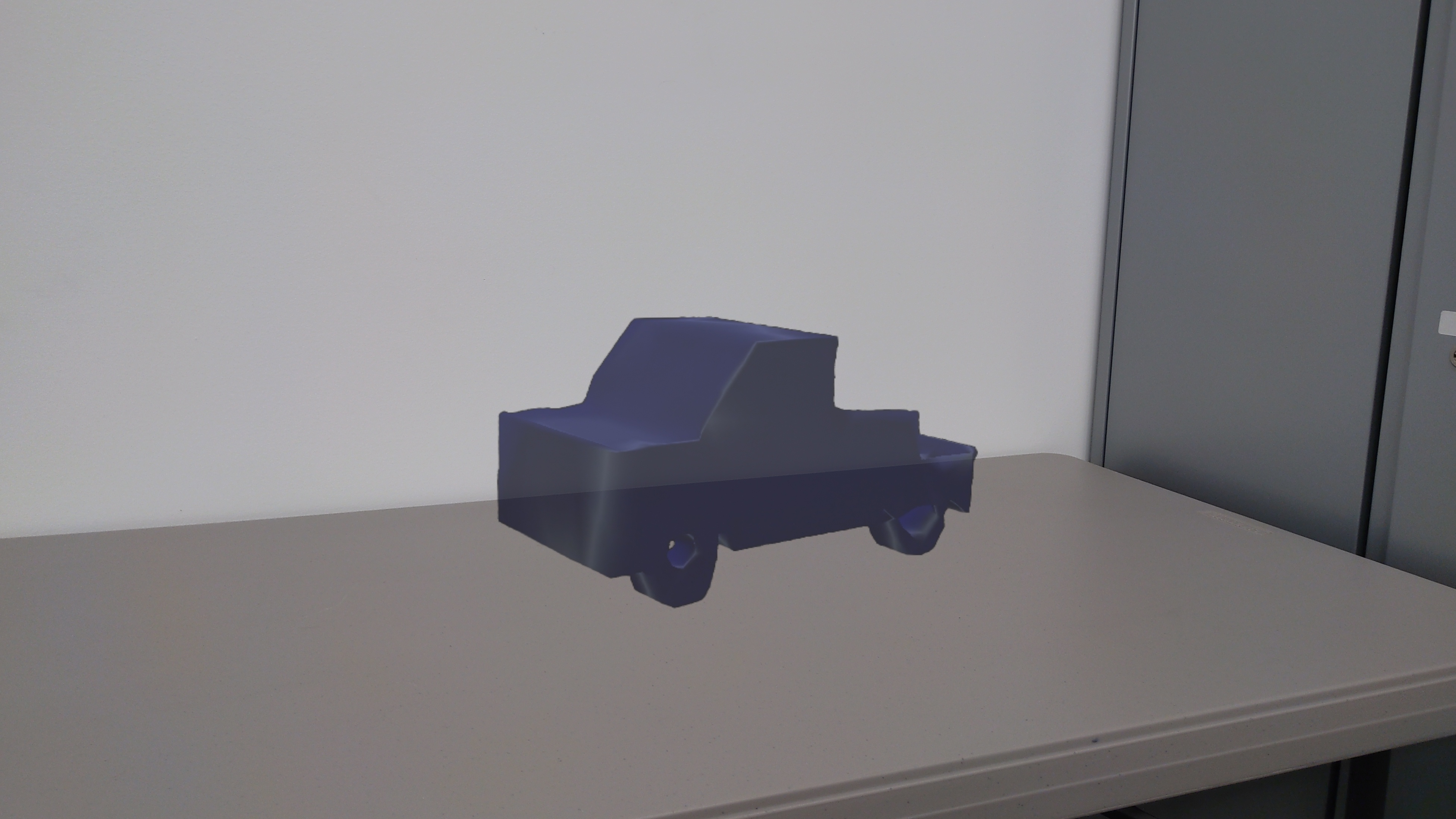}
\includegraphics[width=0.3\textwidth]{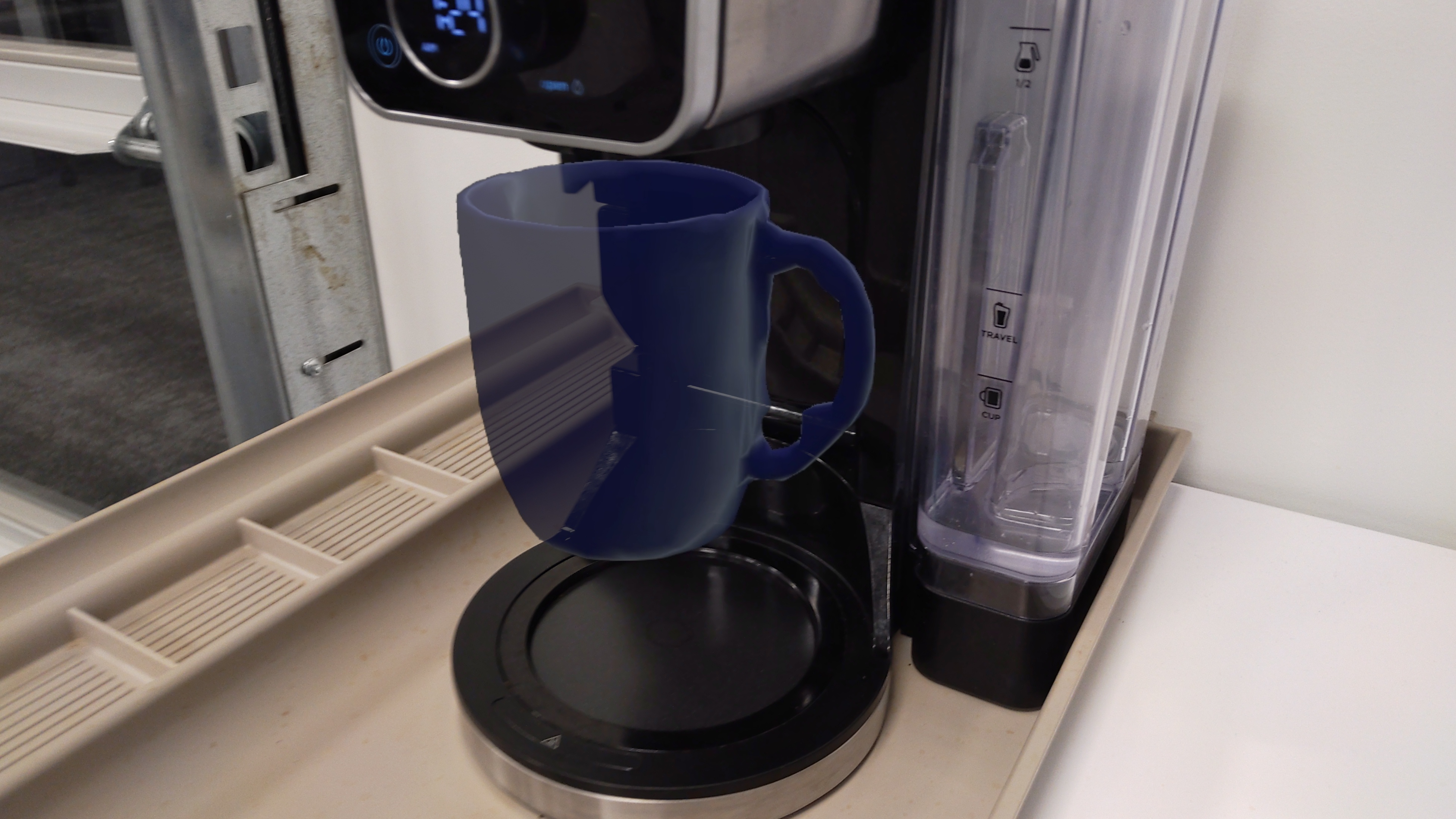}

\caption{Showcasing the transformative power of AI: This sequence of images illustrates \textit{Matrix} AI-powered framework's ability to interactively convert multilingual speech into detailed 3D objects placed in the environment, enriching AR environments and user interactions.}
}

\abstract {
 This paper presents \textit{Matrix}, an advanced AI-powered framework designed for real-time 3D object generation in Augmented Reality (AR) environments.
 By integrating a cutting-edge text-to-3D generative AI model, multilingual speech-to-text translation, and large language models (LLMs), the system enables seamless user interactions through spoken commands.
 The framework processes speech inputs, generates 3D objects, and provides object recommendations based on contextual understanding, enhancing AR experiences.
 A key feature of this framework is its ability to optimize 3D models by reducing mesh complexity, resulting in significantly smaller file sizes and faster processing on resource-constrained AR devices.
 Our approach addresses the challenges of high GPU usage, large model output sizes, and real-time system responsiveness, ensuring a smoother user experience.
 Moreover, the system is equipped with a pre-generated object repository, further reducing GPU load and improving efficiency.
 We demonstrate the practical applications of this framework in various fields such as education, design, and accessibility, and discuss future enhancements including image-to-3D conversion, environmental object detection, and multimodal support.
 The open-source nature of the framework promotes ongoing innovation and its utility across diverse industries.
}


\begin{document}

\firstsection{Introduction}

\maketitle

Generative Artificial Intelligence (AI) is reshaping our world, revolutionizing how we design, learn, and experience technology. Imagine turning simple words into vivid images, lifelike videos, or even interactive 3D objects. These groundbreaking advancements are not just transforming creative industries but are also unlocking new possibilities in education, accessibility, and immersive technologies like extended reality (XR). From empowering educators with dynamic teaching tools to enabling seamless communication across languages and creating accessible environments for diverse users, generative AI is pushing the boundaries of what technology can achieve~\cite{10649569}.

Generative AI have enabling the transformation of natural language into complex outputs, such as 3D objects~\cite{Brown2020}. This capability opens up opportunities for real-time XR applications, particularly in education and accessibility. However, significant challenges persist in realizing the full potential of these technologies. XR solutions for speech-to-3D conversion, such as Dream Mesh~\cite{Weng2024}, which leverages DreamFusion~\cite{poole2023dreamfusion}, suffer from computational latency, often requiring 30–40 minutes to generate a single 3D model, rendering real-time interactions impractical. Furthermore, variability in output consistency, limited multilingual support, and high GPU usage create significant barriers to scalability and inclusivity in XR environments. These limitations highlight the urgent need for innovative approaches to reduce latency, improve output consistency, and enhance overall efficiency.

To bridge this gap, we propose the \textit{Matrix} framework, which accelerates 3D object generation to under 50 seconds using the Shap-E model~\cite{jun2023shape}. \textit{Matrix} addresses AR-specific critical XR challenges by optimizing 3D models for AR-specific devices, reducing latency and computational load while maintaining high responsiveness and inclusivity.

Objectives and Contributions: This paper introduces \textit{Matrix}, a framework designed to enable real-time, multilingual, speech-to-3D object generation in XR applications. It addresses key limitations in latency, consistency, and inclusivity, while also providing an open-source, modular, and adaptable solution for scalable deployment in AR environments.
\textit{Matrix} achieves these objectives by transforming spoken commands into contextually relevant 3D objects. Its multilingual capabilities overcome English-centric limitations in text-to-3D systems, enhancing inclusivity and usability across diverse user groups.

By incorporating multilingual speech-to-text translation using SeamlessM4T~\cite{SeamlessM4T2023} and leveraging LLaMA~\cite{llama2023} for semantic object recognition, \textit{Matrix} enhances user interactions by enabling real-time, context-aware suggestions tailored to individual preferences and histories.

\textit{Matrix} also incorporates semantic search powered by vector databases, which efficiently retrieves pre-generated objects, minimizing GPU usage and addressing inconsistencies in diffusion-based models. This ensures faster response times and optimized resource utilization.

\textit{Matrix} is open-source, designed to run on local systems, enhancing data security with customizable upgrades. It integrates easily with corporate systems, allowing businesses to scale and adapt it to specific needs. Modules for LLMs, text-to-3D, and speech-to-text are interchangeable, ensuring flexibility for various applications while reducing operational costs.
\textit{Matrix} is open-source, designed for local deployment to enhance data security while offering customizable upgrades for scalability. Its modular design allows seamless integration into corporate systems, enabling adaptation to specific application needs. Interchangeable modules for LLMs, text-to-3D, and speech-to-text functionalities ensure flexibility while reducing operational costs.

\textbf{Research Questions:} To guide our exploration, this paper addresses the following questions:
\begin{enumerate}
    \item How can 3D object generation be optimized for real-time AR interactions?
    \item How can GPU usage and model consistency be improved in generative AI systems?
\end{enumerate}

By addressing these questions, this paper provides a comprehensive framework to overcome the existing challenges in AR systems, making them more accessible, responsive, and resource-efficient.

\section{Related Work}

The rapid evolution of AI technologies has enabled significant advancements in interactive systems, particularly in speech-driven interfaces, recommendation systems, and text-to-3D generation. However, despite substantial progress, challenges remain in developing real-time, scalable, and inclusive frameworks for XR environments. This section reviews the most relevant works, focusing on their contributions and limitations.

Speech-to-Text (STT) and Text-to-Speech (TTS) technologies form the foundation of intuitive voice-driven interfaces. SeamlessM4T~\cite{SeamlessM4T2023}, a multimodal translation system supporting over 100 languages, represents a significant step in enabling multilingual and multimodal applications ~\cite{MajidUbiquitous}. WavLLM~\cite{WavLLM2024} further enhances this capability by improving auditory generalization through dual encoders and curriculum learning. These advancements have made STT/TTS indispensable for accessibility, though their integration into real-time XR frameworks remains underexplored.

In addition to STT/TTS, LLMs have demonstrated significant potential for contextual object detection and recommendation. Yan et al.\cite{yan2024ltner} and Ashok and Lipton\cite{ashok2023promptner} utilized GPT-based models to improve Named Entity Recognition (NER) tasks, highlighting LLMs' ability to process complex contexts effectively. For recommendation systems, Kim et al.\cite{kim2024allmrec} and Zhang et al.\cite{zhang2024language} integrated collaborative filtering and domain adaptation into LLM-based recommenders, achieving strong results in text-based scenarios. However, the application of these methods in XR systems, particularly for semantic object retrieval, remains limited.

Generative AI has also transformed text-to-3D technologies, enabling applications such as avatar creation, scene generation, and shape editing. DreamFusion~\cite{poole2023dreamfusion}, for instance, pioneered high-fidelity text-to-3D generation using diffusion models but suffers from long generation times of 30–40 minutes, making it impractical for real-time XR applications. Building on this, Dream Mesh~\cite{Weng2024} incorporated speech-to-3D conversion but inherited similar latency issues. 
Other advancements, such as 
Shap-E~\cite{jun2023shape} improved real-time efficiency by generating detailed and resource-efficient 3D objects.

The \textit{Matrix} framework builds on these advancements, integrating real-time multilingual STT/TTS, LLM-powered semantic object recognition, and Shap-E for optimized text-to-3D generation. Unlike existing systems such as Dream Mesh, \textit{Matrix} addresses key gaps by reducing latency, ensuring consistency, and enhancing scalability for XR environments, making it a robust solution for real-time, accessible, and immersive experiences.

\subsection{Comparison with Similar Works}
The \textit{Matrix} framework for real-time 3D object generation builds upon advancements in generative AI while addressing key limitations in existing systems. Compared to Dream Mesh~\cite{Weng2024}, a Speech-to-3D generative pipeline, \textit{Matrix} demonstrates significant improvements in generation time, consistency, hardware efficiency, and interactivity.

Model Generation Time: Dream Mesh relies on the DreamFusion model, requiring 30–40 minutes for 3D generation~\cite{Weng2024}. \textit{Matrix} reduces this to under 50 seconds by utilizing the Shap-E model and mesh simplification tailored for AR environments.

Model Consistency: Dream Mesh struggles with inconsistencies in identical prompts~\cite{Weng2024}. \textit{Matrix} improves consistency by reusing objects via semantic search in a vector database.

Hardware Requirements: Dream Mesh depends on high-performance GPUs like the Nvidia 3090Ti, while \textit{Matrix} is optimized for mid-range GPUs such as the Nvidia Tesla T4, ensuring broader accessibility.

Additional Features: Unlike Dream Mesh, \textit{Matrix} offers multilingual support, an LLM-based object recommender, and real-time speech feedback, enhancing user interactivity and adaptability for diverse environments.

Table~\ref{table:comparison_framework_dreammesh} summarizes the key distinctions, demonstrating how \textit{Matrix} delivers improved efficiency, consistency, and hardware flexibility for AR applications.

\begin{table}[ht]
\centering
\caption{Comparison of \textit{Matrix} vs. Dream Mesh.}
\renewcommand{\arraystretch}{1.5} 
\begin{tabular}{p{1in}p{1in}p{0.8in}} 
\toprule
\textbf{Feature} & \textbf{\textit{Matrix}} & \textbf{Dream Mesh} \\ 
\midrule
Model Generation Time & \textless{} 50 Seconds & 30-40 minutes \\ 
Model Consistency & Consistent via reuse of objects & Varies with same prompt \\ 
GPU Requirements & Nvidia Tesla T4, 16GB VRAM & Nvidia 3090Ti, 24GB VRAM  \\ 

\bottomrule
\end{tabular}
\label{table:comparison_framework_dreammesh}
\end{table}

The \textit{Matrix} framework incorporates several unique features that enhance usability, inclusivity, and performance. It supports multilingual voice commands, ensuring accessibility for a diverse global user base. Optimized for mid-range GPUs, the framework reduces hardware requirements without compromising performance. By leveraging the Shap-E model and mesh simplification, \textit{Matrix} generates 3D objects in under 50 seconds, enabling real-time interactions. A vector database facilitates semantic object reusability, improving consistency and minimizing computational overhead. Additionally, an LLM-based object recommender and real-time speech feedback enhance interactivity. To further improve the user experience, \textit{Matrix} includes an intuitive user interface with a visual confirmation menu and status board, ensuring transparency, reducing errors, and allowing users to maintain control throughout the interaction process.

\section{Framework Overview and Implementation}

\textit{Matrix} framework functionality is meticulously designed to ensure an intuitive, inclusive, and interactive user experience.
By harnessing state-of-the-art technologies and advanced AI models, we facilitate a seamless transition from verbal commands to visual outputs in real-time, thereby enhancing the capability of users to interact dynamically with the AR environment.
 The system was implemented and tested on a Microsoft HoloLens 2 device, chosen for its advanced spatial computing capabilities and suitability for real-time AR applications. Supporting the HoloLens, we utilized a backend server equipped with an AMD EPYC 7282 16-Core Processor, 32 GB of RAM, and an Nvidia Tesla T4 GPU with 16 GB of VRAM. These specifications allowed us to balance computational performance with real-time responsiveness for AR-specific applications.

\begin{figure}[ht]  
\centering  
\includegraphics[width=\linewidth]{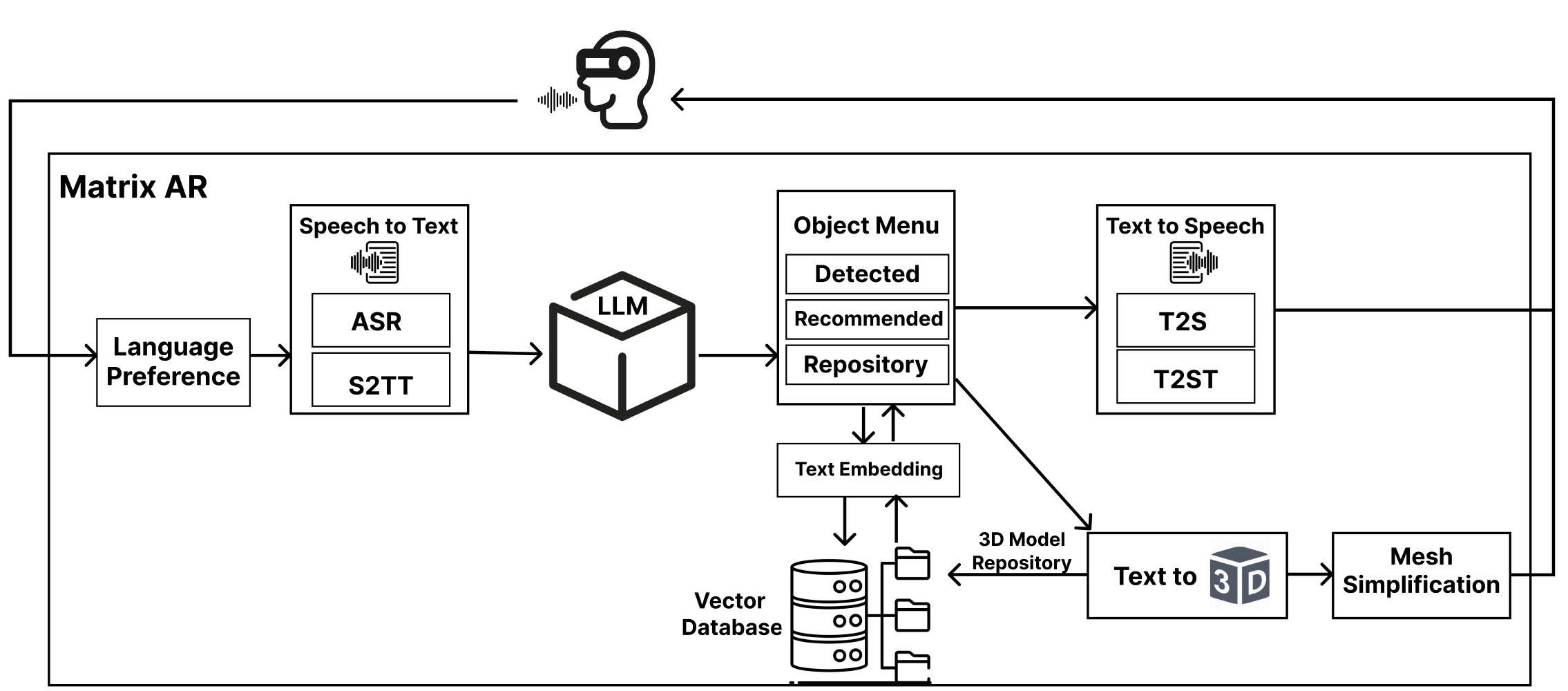}  
\caption{The overview of the developed framework and the individual services provided.}  
\label{fig:Research Framework}  
\end{figure}

We outline the workflow used to interpret user input, generate 3D models, and allow for user-driven customization of the AR environment.
We use an AR application to illustrate the implemented \textit{Matrix} framework service and functionality (\cref{fig:Research Framework}), from initial language selection to the final customization of the interactive environment.
The goal is to maximize user engagement and  responsiveness, ensuring that the technology remains accessible and efficient across various languages and functional demands.
\textit{Matrix} framework \cref{fig:Research Framework}, includes several key components and processes that are labeled with abbreviations to denote specific functions.
S2TT is a Speech-to-Text Translation, indicating the conversion process from spoken language to textual format.
T2S and T2ST denote Text-to-Speech and Text-to-Speech Translation, respectively.
Each component plays a crucial role in ensuring that the functions work seamlessly across various stages of user interaction.

\subsection{Language Selection}

Upon entering the AR environment, users are greeted with a welcoming voice message, initiating a user-friendly introduction to the system.
Following this, they are prompted to select their preferred language from a menu.
This critical step ensures that all subsequent interactions between the user and the system are conducted in the chosen language, thereby tailoring the experience to each individual’s linguistic preference.
The system covers around 37 languages, supported by Meta's SeamlessM4T technology, ensuring inclusivity and accessibility for a diverse user base.

\subsection{Command Initiation}

The AR application utilizes the capabilities of a Microsoft HoloLens 2 device for command voice, setting a high standard for responsiveness and interaction.
To start recording, the application requires the utterance of the wake word, ``MATRIX.''
Upon this command, the application begins recording, and it continues until the user says ``STOP.''
This method allows the user to control when their input is being processed, enhancing privacy and user control over the interaction.

\subsection{Translation and Transcription}
\textit{Matrix} supports interactive communication by detecting user speech and responding in their input language. Non-English speech is translated into English using the SeamlessM4T model, ensuring compatibility with diverse languages. English inputs are directly transcribed by SeamlessM4T’s ASR capabilities, optimized for high accuracy.
If the user’s language is not English, the system translates the output text back into their language via SeamlessM4T, ensuring accessibility. Spoken responses in English are generated using Coqui.ai TTS, delivering natural, engaging interactions.

The system processes audio commands up to 15 seconds long, ensuring real-time response and avoiding delays, providing an efficient AR user experience.

\subsection{Object Extraction}

For extracting meaningful content from the transcribed or translated text, we employ the Llama2 7B LLM, which identifies objects and their attributes within the user’s commands.
To this end, we use an optimized version of Llama, llama.cpp, designed to run with minimal hardware.
This enables \textit{Matrix} framework to be used on a wide variety of hardware locally, ensuring broader accessibility and flexibility.
The system identifies object properties such as size, shape, and color, which are essential for dynamic user interaction. The langchain library further refines the prompts for extracting highly specific object attributes, ensuring relevance and precision.

\subsection{Object Recommendation}

Once the objects are identified and modeled from the transcribed or translated texts, the next step is to enhance user interaction by suggesting related objects.
This is achieved through a sophisticated use of the Llama2 LLM, which processes the extracted object data to generate recommendations.
The LLM considers the context, properties, and functions of the identified objects to suggest objects that could logically coexist or complement the primary objects in a given environment.
This process not only enriches the user experience by providing creative and contextually appropriate options but also aids in developing a more interactive and immersive AR environment.

\subsection{Search Repository}
 The application integrates a 3D object repository that facilitates efficient retrieval of matches or similar items based on user requests. This repository, containing pre-generated models and past user interactions, plays a critical role in minimizing GPU usage and computational demand while supporting diverse AR scenarios. However, the repository has limitations in handling highly unique or custom user requests, which may require on-the-fly generation.

A vector database optimizes the search process by identifying contextually similar objects through pre-embedded vectors. This approach minimizes duplicate suggestions and enhances user interaction by efficiently retrieving relevant objects based on the context of the user's input. Additionally, the vector database reduces system resource requirements, ensuring a balance between performance and responsiveness.

The combination of LLM-generated recommendations with repository searches provides users with a comprehensive range of options, accelerating object selection and improving interaction quality. The repository is continuously updated to maintain relevance, supporting a dynamic, adaptive, and efficient system for real-time object integration in AR.

\subsection{Object Generation and Mesh Simplification}

The extracted objects and related items from the LLM and repository are presented to users via an interactive menu, accompanied by auditory prompts to guide object selection for the AR space. Repository objects load directly into the AR environment without additional GPU use, reducing computational demand and improving responsiveness.

For other menu selections, text is converted into 3D models using Shap-E, with parameters like 64 Karras-inspired sampling steps, a sigma range of 1e-3 to 160, and s\_churn=0 for stable denoising. Once generated, the model is displayed, and a voice message confirms its creation in the user's chosen language.

Selected objects are embedded using the all-MiniLM-L6-v2 model and stored in the Chroma vector database. This enables efficient suggestions of related objects in future interactions, enhancing the user experience by streamlining access to relevant items and improving AR interactivity.

After generating the 3D model, the system applies a mesh simplification process using the Quadric Edge Collapse Decimation algorithm. This process minimizes the number of vertices and faces, effectively reducing the model’s complexity while preserving its essential structure and visual fidelity. By optimizing the model to approximately 1,000 vertices after performing five iterations of simplification, the framework ensures a balance between visual detail and performance, creating lightweight models that integrate seamlessly into AR environments. The default Shape-E model outputs average 13,944 vertices and 27,884 faces, offering high detail but being too large for real-time use. Reducing the number of vertices reducing model size and improving performance.

\subsection{Interactive Environment Customization}

Following the placement of each object, users can interact with them using hand gestures.

\begin{figure}[ht]  
\centering  
\includegraphics[width=\linewidth]{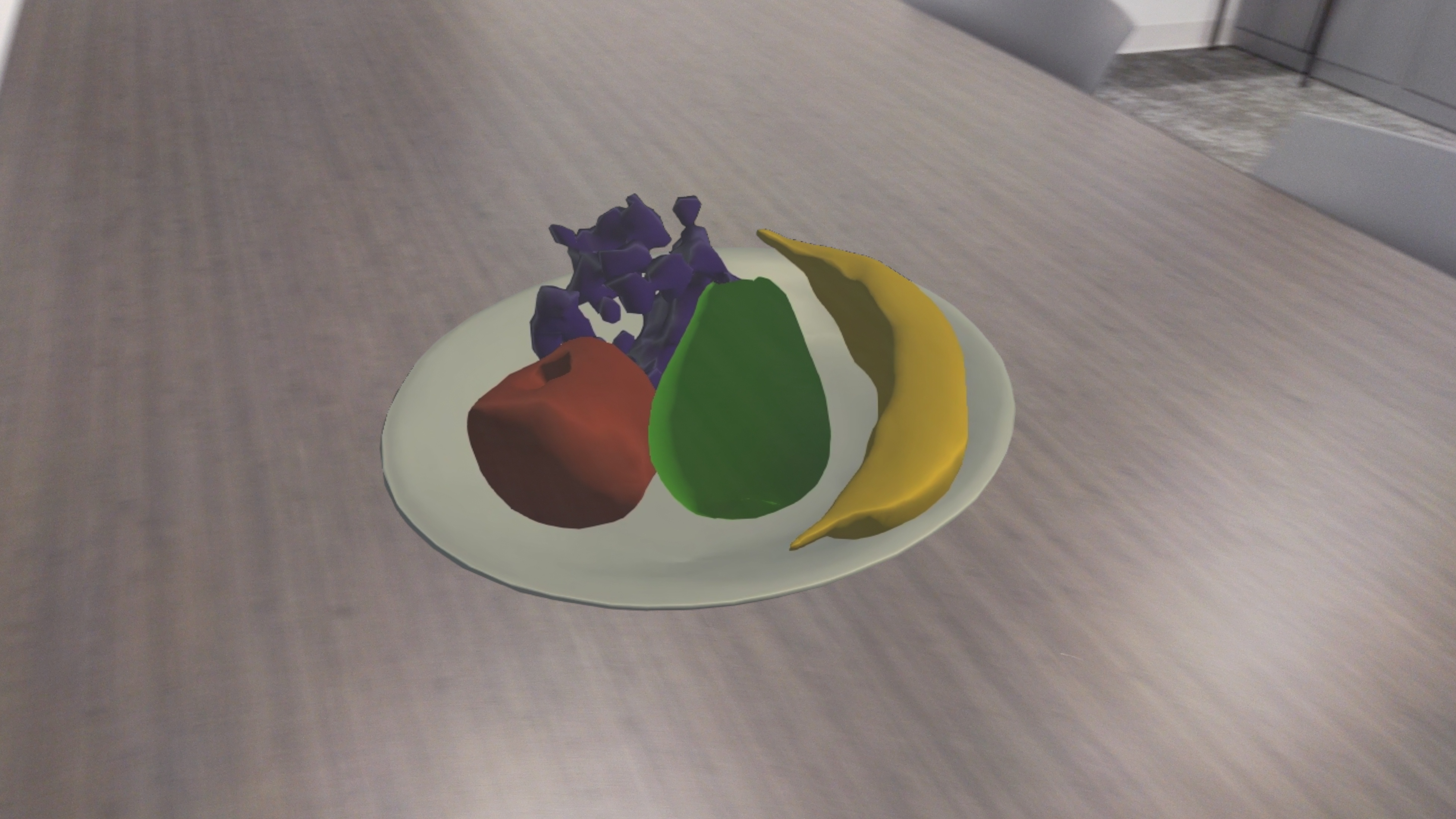}  
\caption{A plate containing a banana, a red apple, a pear, and grapes, generated in real time using \textit{Matrix}, is placed on a table by the user. The objects were created based on five voice commands given by the user: “Matrix, create a banana,” “Matrix, create a red apple,” “Matrix, create a pear,” “Matrix, create grapes,” and “Matrix, create a white plate.” The plate was retrieved from the repository, while the other objects were generated using text-to-3D conversion.}
\label{fig:fruit}  
\end{figure}

Hand gestures allow users to effortlessly move, resize, and rotate each item, enabling them to customize their AR spaces to their liking. Objects can be moved by pinching the center of the object and dragging the hand, resized by pinching two corners of the object with both hands, and rotated by pinching the center of the object and twisting the wrist.
Designing a fruit arrangement in AR, as shown in \cref{fig:fruit}, exemplifies this interactive customization process.

\begin{figure}[ht]  
\centering  
\includegraphics[width=\linewidth]{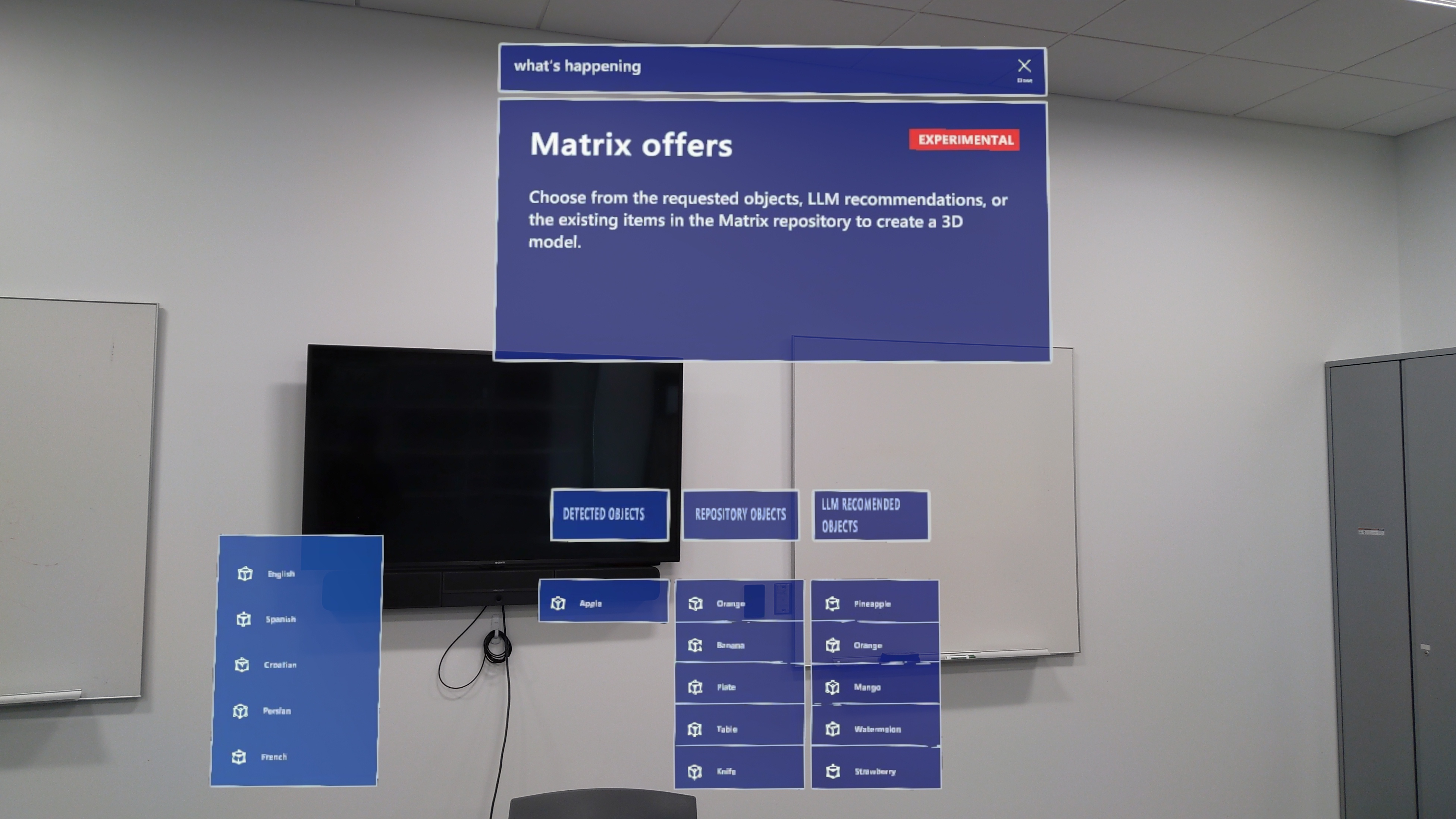}  
\caption{
Status board example in \textit{Matrix} AR application.
The ``what's happening'' board displays the ``Matrix offers'' message, informing users to choose from requested objects, LLM recommendations, or items from the repository to create a 3D model.
}  
\label{fig:status}  
\end{figure}

Finally, to ensure optimal user experience and system transparency, we have integrated a status board at the top of the user interface, as seen in \cref{fig:status}.
This board continuously displays real-time updates about the system's status, showing messages such as `Welcome,' `Listening,' `Thinking,' `Offers,' `Baking,' and `Presenting,' depending on the current state of the system when processing-intensive tasks are being executed.

\section{User Study and Performance Evaluation} 

This research was conducted under the oversight of the Virginia Tech Institutional Review Board (IRB) (IRB Protocol Number: 24-1072). The protocol, titled ``AI-Powered 3D Model Generation and Recommendation Systems in Augmented Reality,'' was determined to meet the criteria for exemption under 45 CFR 46.104(d) categories 2(ii) and 3(i)(B). As part of this exemption, all activities were reviewed and deemed compliant with ethical standards for research involving human subjects.

To comprehensively assess the effectiveness and performance of \textit{Matrix}  framework for real-time 3D object generation in AR environments, we defined key performance metrics that spanned system performance.
These metrics allowed us to evaluate the technical efficiency of the framework.
To validate these metrics, we conducted a user study with 35 participants from diverse backgrounds, with ages ranging from 18 to 40, ensuring insights from a broad generational spectrum. Gender representation included 63\% male, 37\% female, and provisions for inclusivity in gender identification. Participants displayed varied familiarity with AR/VR, with most reporting rare to occasional usage, highlighting a mix of novice and moderate users. Experience with 3D design tools showed a majority slightly familiar (54\%), while comfort levels with AR headsets varied, with 34\% moderately comfortable and 17\% having no prior usage. This diverse demographic provided valuable insights into usability across different experience and comfort levels with AR/VR and 3D design tools.
The evaluation was split into multiple phases, focusing on task performance, GPU utilization, system resource consumption, and model accuracy.

\subsection{Tasks}

Participants were asked to complete a set of tasks that highlighted various features of \textit{Matrix}.
Each participant completed all three tasks sequentially, ensuring a comprehensive evaluation of the system's core functionalities, interaction capabilities, and performance.
The tasks included:

\begin{description}
\item[Task 1:]
Generate a simple 3D object using a voice command.
This task was meant to test the core functionality of the speech-to-3D conversion process, including how well the system transcribes the user's speech and generates an object based on that input.
\item[Task 2:]
Retrieve pre-generated objects from the repository.
This task was designed to evaluate how efficiently the system retrieves objects from the repository, and whether this retrieval is faster than generating objects from scratch.
\item[Task 3:]
Customize object placement and size in the AR environment.
This task required participants to manipulate the generated objects in real-time, testing the interaction capabilities of the system, including how intuitive and responsive the object manipulation felt to the users.
\end{description}

\section{Results}

We now discuss user study findings followed by performance evaluation results. 

\subsection{Evaluation of System Usability Scale}

The System Usability Scale (SUS) assessed user perceptions of system usability, with scores adjusted per SUS guidelines and scaled to 0-100 \cite{brooke1996quick}. The average SUS score was \textbf{69.64}, slightly above the usability threshold of 68, indicating general usability with areas for improvement. Key findings include strengths in ease of learning and frequent use, with users reporting confidence in using the system. Challenges were noted in complexity and the need for support in certain features.


An Analysis of Variance (ANOVA)  revealed a significant difference in SUS scores between user groups (\(\textit{F} = 18.21, \textit{p} < 0.001 \)).The Rarely/Never group, consisting of users who used XR system less than 3 times total, had a lower mean score (\textbf{64.38}) compared to the Regularly/Sometimes/Often group, which includes users who interact with the XR system 3 or more times total, (\textbf{80.71}). This highlights how frequency of use influences perceptions of usability.

In summary, the SUS score indicates a usable system with strong user confidence but areas for refinement in complexity and support. Future iterations could streamline features and promote regular use to further improve usability.

\subsection{Qualitative User Study Results}

Participants appreciated features like intuitive voice commands, touch gestures, and the ability to manipulate objects (e.g., moving, resizing, rotating) seamlessly. The system’s capability to scan environments and recommend objects contextually was highly valued, especially for professional applications like interior design and prototyping. However, challenges included latency in 3D model generation, inconsistent rendering quality, and difficulties with interface navigation and object alignment. Suggested improvements included enhanced 3D model realism, reduced latency, simplified navigation, and improved depth alignment.

\subsection{Task Performance Evaluation}

We evaluated task performance using five metrics: task completion time, success rate, error rate, system responsiveness, and mesh size.

We evaluated task performance using five metrics: task completion time, success rate, error rate, system responsiveness, and mesh file size. Task completion time was measured as the average time to complete tasks, which was significantly reduced by AR’s real-time guidance~\cite{RajSub2023}. Task success rate referred to the frequency of successful first attempts and was impacted by factors such as speech and object manipulation errors~\cite{ErrorRate2013}. System error rate, which measured the frequency of errors like incorrect object generation, was used as an indicator of system robustness~\cite{ErrorRate2013}. System responsiveness, defined as the time from input to rendering, was critical for ensuring smooth interaction~\cite{Bi2023MISARAM}. Finally, mesh file size was optimized through vertex reduction techniques, which enhanced AR performance~\cite{Siddeq2016}.

Task completion time was measured as the average time to complete tasks, which was significantly reduced by AR’s real-time guidance~\cite{RajSub2023}. Task success rate referred to the frequency of successful first attempts, defined as the percentage of tasks completed correctly on the initial try without requiring retries or corrections. It was impacted by factors such as speech recognition errors (e.g., incorrect transcription of voice commands) and object manipulation errors (e.g., difficulty in accurately positioning, resizing, or rotating objects) ~\cite{ErrorRate2013}. System error rate, which measured the frequency of errors like incorrect object generation, was used as an indicator of system robustness~\cite{ErrorRate2013}.  System responsiveness, defined as the time from input to rendering, was critical for ensuring smooth interaction~\cite{Bi2023MISARAM}.  Finally, mesh file size was optimized through vertex reduction techniques, which enhanced AR performance~\cite{Siddeq2016}.

\Cref{table:task_performance_reduced_mesh} summarizes performance metrics, showing Task Completion Time dropping from 386s (default) to 122.4s (optimized), demonstrating the importance of file size reduction for real-time AR.


\begin{table}[ht]
\centering
\caption{Task performance with default Shap-E model output and reduced mesh size.}
\begin{tabular}{p{1.8in}p{0.5in}p{0.5in}} 
\toprule
\textbf{Metric} & \textbf{Shap-E Output} & \textbf{Reduced Size} \\ 
\midrule
Task Completion Time (seconds) &  386.2  & 122.4  \\ 
Task Success Rate (\%) & 76\% & 88\% \\ 
System Error Rate (errors/task) & 0.15  & 0.14  \\ 
System Responsiveness (seconds) & 282.73  & 74.32  \\ 
Mesh File Size (MB) & 2.5  & 0.16  \\ 
\bottomrule
\end{tabular}
\label{table:task_performance_reduced_mesh}
\end{table}

\subsection{Model Accuracy}

Finally, we measured the accuracy of the AI models used in \textit{Matrix}, focusing on speech recognition and object generation. Accurate model performance is essential for delivering a smooth and error-free user experience. \cref{table:model_accuracy} summarizes these accuracy metrics.
 
Speech-to-text accuracy evaluates the rate at which the system correctly converts spoken commands into text \cite{gyulyustan2024measuring}. In the user study, we tested the system's performance specifically for English commands, assessing its reliability in accurately transcribing spoken inputs.  Object recognition precision measures how accurately the system identifies and suggests relevant objects based on the transcribed text \cite{Obj2025}. Object rendering fidelity measures how accurately generated 3D objects match the user's spoken command.\cite{cheng2023dnarender}. 

\begin{table}[ht]
\centering
\caption{Summary of AI model accuracy metrics for speech recognition, object recognition with LLM and object generation.}
\begin{tabular}{lcc} 
\toprule
\textbf{Metric} & \textbf{Value} \\ 
\midrule
Speech-to-Text Accuracy (\%) & 91\% \\ 
Object Recognition Precision (\%) & 85\% \\ 
Object Rendering Fidelity (\%) & 87\% \\ 
\bottomrule
\end{tabular}
\label{table:model_accuracy}
\end{table}

\section{Discussion}
This section addresses the research questions outlined in the introduction, providing insights into the implications of optimizing 3D object generation for real-time AR interactions and improving GPU usage and model consistency in generative AI systems.

\textit{Matrix} demonstrates significant potential for optimizing real-time 3D object generation in AR by reducing task completion times and improving system responsiveness.

In education, \textit{Matrix}  bridges auditory and visual reinforcement by enabling students to verbalize words and see corresponding 3D objects. Students can verbalize words in a new language and see corresponding 3D objects, improving vocabulary retention and pronunciation. Teachers can design interactive lessons where students use speech to interact with 3D objects, creating engaging learning experiences. This aligns with research showing how multimodal learning enhances retention and engagement~\cite{Sarshartehrani2024}.

Optimizing GPU usage is critical for real-time AR applications. The pre-generated object repository implemented in  \textit{Matrix} reduced GPU utilization by minimizing the need for on-the-fly generation. Additionally, employing semantic search through a vector database enabled object reuse, enhancing model consistency by reducing output variability—a key limitation in diffusion-based models~\cite{poole2023dreamfusion}.

For instance, a designer could describe a concept, view it in 3D, and refine it through iterative adjustments, promoting creativity and efficiency. The integration of scalable AR platforms to address challenges like inconsistent data acquisition~\cite{ganj2023} highlights the practical applications of our framework.

The \textit{Matrix} enhances accessibility by enabling visually impaired users to engage with digital environments using tactile 3D models and audio feedback. This inclusive approach, combined with speech recognition and object generation, ensures meaningful interaction with AR systems for all users, particularly in education.

\section{Conclusion and Future Work}
 
The work presented in this paper lays a foundation for integrating advanced generative AI models into AR systems, demonstrating significant improvements in technical efficiency.
Through optimization of 3D object generation, system responsiveness, and resource management, we have developed \textit{Matrix}, a framework capable of addressing the critical challenges posed by AR environments.
This includes reducing the file sizes of 3D models, improving system performance on resource-constrained devices, and enabling real-time object generation that enhances user engagement.

We aim to enhance the Matrix by integrating voice-to-3D model conversion, alongside image-to-3D transformation features, enabling seamless creation of 3D models from both audio and visual inputs ~\cite{behravan2025HCII}. Additionally, by incorporating Vision-Language Models (VLM), the system will analyze environments in real-time to suggest or generate 3D objects, fostering dynamic interactions between real-world and virtual elements within the Matrix ~\cite{BehravanVRST,behravan2025AIxVR}.

Despite its many strengths, the \textit{Matrix} framework has some limitations that must be addressed in future work. The object generation process occasionally struggles with abstract or highly complex descriptions, resulting in objects that may not align with user expectations. Although optimized for real-time interactions, latency can still occur when generating 3D models on resource-constrained devices, potentially impacting user experience in highly interactive applications.

\bibliographystyle{abbrv-doi}
\bibliography{IEEEVR_2025_3d}

\end{document}